# Broadband transverse susceptibility in multiferroic Y-type hexaferrite $Ba_{0.5}Sr_{1.5}Co_2Fe_2O_{22}$

P. Hernández-Gómez*, D. Martín-González, C. Torres, J. M. Muñoz

Dpt. Electricidad y Electrónica, Universidad de Valladolid, Paseo de Belén 7, 47011 Valladolid Spain.

*Corresponding author

e-mail: pabloher@ee.uva.es

Tel: +34983423895

## ABSTRACT

Single phase multiferroics in which ordered magnetic and ferroelectricity coexist, are of great interest for new multifunctional devices, and among them Y-type hexaferrites are good candidates. Transverse susceptibility measurements, which have been proved to be a versatile tool to study singular properties of bulk and nanoparticle magnetic systems, have been carried out with a broadband system on polycrystalline Y type hexaferrites with composition $Ba_{0.5}Sr_{1.5}Co_2Fe_2O_{22}$, optimal to exhibit multiferroic properties. In the temperature range 80-350 K transverse susceptibility measurements with DC fields up to ±5000 Oe reveal different behaviour depending on the sintering temperature. The thermal evolution of the anisotropy field peak exhibit four regions with different slopes: positive in 80-130 K, negative in 130-200 K, constant in 200-280 K and negative in 280-350 K, which can be considered a signature of spin transitions in this compound.

## Introduction

The study of multiferroic materials in which ordered magnetic and ferroelectricity coexist by means of the magnetoelectric effect, has been the subject of intensive study in recent years due to the potential technological applications in which magnetic properties are controlled with electric fields and viceversa, like spintronic devices, magnetoelectric sensors, electronic devices, storage applications and medical drug delivery [1-4]. In particular, single phase multiferroics are of great interest for these multifunctional devices. Among the compounds with this characteristic, ferroxplana type hexaferrites with Z ($[BaSr]_3Me_2Fe_{24}O_{41}$) and Y ($[BaSr]_2Me_2Fe_2O_{22}$) phases are very promising materials, due to the giant magnetoelectric coupling between magnetism and ferroelectricity and low magnetic fields to switch the electric polarization [5, 6]. In Z phase this behaviour is observed at room temperature, but single Z phase formation is still challenging. The magnetoelectric coupling and temperature range in Y-type hexaferrite can be improved by appropriate cation substitution and varying sintering conditions [7-10]. The crystal structure of Y-type hexaferrite belongs to the R-3m space group with hexagonal structure, and can be obtained with the piling up of S ($Me_2Fe_4O_8$) and T ($BaFe_8O_{14}$) blocks along the c axis. Metal cations occupy four octahedral ($18h_{VI}$, $6c_{VI}$, $3a_{VI}$ and $3b_{VI}$) and two tetrahedral sites ($6c_{IV}$ and $6c_{IV}*$). Regarding the magnetic structure, the planes containing Ba and Sr ions have small net magnetic moments, whereas the rest of layers have large magnetic moments, so that we can consider them as S and L blocks [6], with a net moment perpendicular to the c axis. The substitution of Sr replacing Ba causes lattice deformation and hence the superexchange bond angles increase, especially the corresponding to the metal cations in the layers of Ba(Sr) ions. Due to this deformation the compound can exhibit different magnetic structures i.e. proper screw, longitudinal conical (LC), transverse conical (TC), double fan (FE) or collinear, and some of them allow the existence of an induced electric polarization that can be explained according to the inverse Dzyaloshinskii-Moriya effect or spin current model [5].

Transverse susceptibility (TS) is obtained when applying a bias DC magnetic field, and AC applied field and the magnetic response is measured in a transverse direction. It has been proved to be a versatile tool to study singular properties of bulk and nanoparticle magnetic systems, especially to obtain their anisotropy and switching field [11, 12], and is also a probe of phase transitions caused by anisotropy [13]. Usually it is measured with the help of a self-resonant LC circuit, with high sensitivity but with frequency limitations. We have developed a fully automated, broadband system based on a LCR, that allows this measurement in varying ranges of DC and AC applied fields, temperature and frequency with enhanced sensitivity. TS measurements have been carried out on Y type hexaferrites with composition $Ba_{0.5}Sr_{1.5}Co_2Fe_2O_{22}$, (BSCFO) optimal to exhibit multiferroic properties [14, 15]. In addition, Co substitution can enhance the magnetoelectronic effect in hexaferrites [16]. It is expected that the different spin configurations modify the magnetic properties detected with TS measurement, so that it can be a versatile tool to stablish the different temperature regions in which the magnetoelectric effect occur.

**Materials and methods**

Polycrystalline samples have been prepared by means of standard ceramic techniques. Stoichiometric amounts of $SrCO_3$ (98%), $BaCO_3$ (99%), $Co_3O_4$ (98%), and $Fe_2O_3$ (99%) with molar proportions 4.5: 1.5: 2: 18, corresponding to the initial composition of Y type hexaferrite $Ba_{0.5}Sr_{1.5}Co_2Fe_{12}O_{22}$ (BSFCO), were mixed in agate mortar, calcined at 900 ºC to remove carbonates, pressed in the form of rods with 5 mm of diameter and 20 mm in length, and then sintered in air atmosphere at 1050 ºC, 1150 ºC and 1250 ºC, according to the phase diagrams [17].

X ray diffraction patterns were obtained with a diffractometer Bruker Discover 8 at room temperature employing Cu-Kα radiation (λ=1.54056 Å). Intensity data were collected by the step-counting method (step 0.02 º/s) in the range 20º < 2θ < 70º. Quasi static hysteresis loops were obtained in powdered samples with an inductive technique at room temperature with a maximum field of 4500 Oe. Homemade control program also corrects for the shape demagnetization factor.

Transverse susceptibility measurement were carried out with a broadband automatic system in which the sample rods form the core of a coil that produces an longitudinal AC magnetic field with frequency 1 kHz and maximum amplitude of 2 Oe. The sampleholder with the coil and a heating element is put inside a cryostat to allow temperature measurements in the range 80 K- 350 K. The cryostat tail lies into the polar pieces of an electromagnet fed by a power supply Agilent 6675A that produces a DC magnetic field, measured with a FW Bell 5080 gaussmeter, perpendicular to the AC magnetic field. DC bias field sweeps run from +5000 Oe to -5000 Oe, and then sweep back to +5000 Oe (i.e. bipolar scans), according to measurements found in literature [18]. The response of the measuring coil is obtained with an LCR meter Agilent E4980A. Temperature control is achieved with a data logger Hewlett Packard 3497A. All the system is controlled via GPIB with a PC by means of a homemade control program with Agilent VEE software. The broadband nature of our system arises from the possibility of varying temperature, DC magnetic field, AC field frequency and amplitude with enhanced sensitivity, overcoming the frequency limitations of resonant circuits.

**Results and discussion**

In Figure 1, a representative XRD pattern of Y type hexaferrite sample sintered at 1050 ºC is presented together with the pattern corresponding to the JCPDS card 40-1047. The diffraction pattern reveals that this sample is a single phase hexagonal structure with space group R-3m, as it also happen with the sample annealed at 1150 ºC [15]. Sample sintered at 1250 ºC corresponds to Y type hexaferrite with secondary phase of Z type. The lattice parameters obtained for these samples reveal minor changes with the increase in sintering temperature: a= 5.8447 Å, 5.8477 Å and 5.8478 Å and c= 43.4079 Å, 43.3658 Å and 43.4414 Å, in good agreement with literature data [15].

Hysteresis loops of the as calcined and sintered Y type hexaferrites are shown in Figure 2. The calcined sample exhibits a rather wide loop, as the sample is composed mainly by M-type hexaferrite, according to phase diagrams [17]. Y type formation is achieved by sintering at 1050 ºC, and the hysteresis loop is notably narrower, according to the soft magnetic character of ferroxplana compounds. Higher sintering temperature promotes higher crystallinity and densification, resulting in higher magnetization and smaller coercive fields: 157.5 Oe, 102 Oe and 59 Oe for 1050 ºC, 1150 ºC and 1250 ºC resp.. The use of the law of approach to saturation can give us a reference value of saturation magnetization in these compounds: 20.4 emu/g, 21.6 emu/g and 33.2 emu/g for 1050 ºC- 1250 ºC sintered samples (data have been converted to allow comparison with literature data). We can observe a clear increase in saturation magnetization in the sample sintered at 1250 ºC, caused by the higher crystallinity and also by the increased contribution of the secondary Z phase.

The transverse susceptibility ratio can be expressed as [18, 19]

$$\frac{\Delta\chi_T}{\chi_T}(\%) = \frac{\chi_T(H) - \chi_T(H_{SAT})}{\chi_T(H_{SAT})} \times 100 \quad (1)$$

where $\chi_T(H_{SAT})$ is the transverse susceptibility at the saturating field ($H_{SAT}$ =5000 Oe in our case). For samples filling completely the measuring coil we can deduce [18]

$$\frac{\Delta\chi_T}{\chi_T}(\%) \approx \frac{L(H) - L(H_{SAT})}{L(H_{SAT})} \times 100 \quad (2)$$

Since this is a measure of overall change in transverse susceptibility, there is no dependence on the geometrical parameters, and then it is useful to extract several parameters, as transition temperatures, anisotropy and switching fields in many systems [18, 19]. When $\Delta\chi_T/\chi_T$ is represented as a function of DC field, maxima are observed at $H_{DC}=\pm H_A$ and effective anisotropy constant can be deduced. The expected maximum at $H_{DC}=H_s$. i.e. the switching field, is often merged to one of the former in systems with distribution of anisotropy fields.

In the Figure 3 a representative 3D plot of bipolar broadband TS measurements in the temperature range 80-350 K and DC fields up to ±4000 Oe is shown (frequency and AC field amplitude are kept constant at 1 kHz and 2 Oe) for the sample sintered at 1050 ºC. We can observe that at low measuring temperatures the amplitude of TS is small and increases strongly over room temperature. This is a general trend, but the increase is not so remarkable in the sample annealed at 1150 ºC, and the increase is gradual in the sample sintered at 1250 ºC. The maximum TS value takes place at the top measuring $T_{meas}$ tested, and diminishes from 8% for the sample sintered at 1050 ºC, to 5.5% and 3% for samples prepared at 1150 ºC and 1250 ºC resp. We can then conclude that the higher the crystallite size, the lower the TS. It is known that the main contribution to the TS is provided by the grains with easy axis oriented perpendicular to the DC field [11], so that bigger grains reduce the amount of material with the optimal orientation measured by TS.

Deeper insights are obtained with the 2D plots of Figure 4, in which the DC field effect is highlighted. We can observe two different behaviours in all the samples with the measuring temperature: at higher $T_{meas}$ we can observe only one peak, while at low $T_{meas}$ it can be seen that a bipolar scan starting from $+H_{SAT}$ exhibit a maximum at $+H_A$, then in the sweep from 0 to $-H_{SAT}$ a minimum and a smaller maximum is observed, or a maximum with a shoulder. The sweep from negative to positive $H_{SAT}$ is similar. Different heights of maxima in the two bias field sweeps are indicative of temperature variations of the sample or thermal relaxations, and are absent in this study, whereas different height in the two maxima in a unipolar sweep are indicative of interparticle interaction as well as anisotropy field dispersion [19]. Peaks are all rounded caused by the expected distribution of anisotropy fields according to the grain size distribution in polycrystals [11], that in addition cause the observed shoulder in the TS curve [12], so that we have only considered the higher maxima. Comparing the results we can observe strong differences in the behaviour depending on the sintering temperature. In the Figure 4a we show the $\Delta\chi_T$ corresponding to the sample sintered at 1050 ºC, in which the maximum corresponding to $H_A$ shifts to smaller absolute values with increasing measuring T. Sample fired at 1150 ºC behaves in a similar manner but the peaks are wider, according to the higher grain size distribution. This behaviour is more evident in the sample sintered at 1250 ºC (Figure 4b).

However, the most striking feature in this sample is that the peak initially shifts to higher values of $H_A$ and then diminishes. The temperature dependence of the anisotropy fields obtained as mentioned above from TS measurements is represented in the Figure 5. We observe that in the sample prepared at 1050 ºC, composed by single phase Y-type hexaferrite, the anisotropy field is almost constant (680 Oe) until $T_{meas}$= 280 K, then diminishes. On the other hand, sample annealed at 1250 ºC exhibit four regions with different slopes: positive in 80-130 K, in which the anisotropy field increases from 800 to 1400 Oe, then negative in 130-200 K down to 680 Oe, constant in 200-280 K and negative in 280-350 K, with values in this region very similar to the previous sample. Finally, sample fired at 1150 ºC exhibit a behaviour in between the others, with a small increase and decrease of anisotropy field at low $T_{meas}$ (110 and 160 K), and a similar behaviour over 200 K, with a higher anisotropy field between 200 and 280 K (900 Oe), and identical slope in the range 280 to 350 K.

From neutron diffraction studies in very similar composition [14] it is known that the magnetic structure of BSCFO at zero field is alternate longitudinal conical (ALC) among the L and S blocks from 3 K to 280 K, among 280 K and 360 K the magnetic order changes to a mixed conical or double fan structure (FE), and then exhibits a proper screw structure. It is also known that the magnetic structures are different before and after application of external magnetic field, when the system undergoes field induced magnetic phase transitions, and at different temperatures the system do not return back to the previous magnetic structure [9]. In addition, the magnetic field required to induce a magnetic phase transition changes depending on the metallic cations employed. In $Ba_{0.5}Sr_{1.5}Co_2Fe_2O_{22}$ polycrystalline samples [15] the ZFC magnetization exhibits a strong change among 120 K and 270 K, and a cusp in FC at 200 K. With this in mind, we can consider the thermal evolution of anisotropy field obtained with TS measurements a signature of the spin transitions in these compounds.

According to our results, we consider three different regions: 80-200 K, 200-280 K and 280-350 K. The latter corresponds to ALC magnetic phase and is present in all our samples. The region among 200 and 280 K exhibits mixed properties between ALC for grains in which H is parallel to c axis, and FE for grains with c perpendicular to H [20]. Application of magnetic field promote a magnetic transition to FE that persist when the field is removed [9] and produce the parameters measured in TS. In this case the increase in sintering temperature have small effects mainly due to the different anisotropy arising from the increasing grain size. Finally, in the region 80-200 K the sintering temperature and bias field can induce several changes. For the 1050 ºC sample the bias field applied is not strong enough to induce the magnetic transition and remains in ALC state, whereas for 1250 ºC the system undergoes the magnetic transition to FE [9] and the strong variation in the $H_A$ detected with TS is caused by the substantial change of magnetization in this temperature range [9, 15] Finally, sample fired at 1150 ºC probably achieve an intermediate state, so that the effects observed at higher sintering temperatures are only envisaged.

## Acknowledgements

Funding: This work was supported by the Spanish Ministerio de Ciencia Innovación y Universidades, (AEI with FEDER), project id. MAT2016-80784-P

**Figure Captions**

**Figure 1**. X ray diffractogram of the sintered $Ba_{0.5}Sr_{1.5}Co_2Fe_2O_{22}$ hexaferrite sample at 1050º C, and the corresponding JCPDS card.

**Figure 2**. Hysteresis loops of the $Ba_{0.5}Sr_{1.5}Co_2Fe_2O_{22}$ hexaferrite calcinated at 900 ºC and the sintered samples at 1050º C, 1150º C, and 1250º C.

**Figure 3**. 3D view of the broadband transverse susceptibility of the $Ba_{0.5}Sr_{1.5}Co_2Fe_2O_{22}$ hexaferrite sample sintered at 1050º C.

**Figure 4**. 2D plot of transverse susceptibility vs DC magnetic field of the $Ba_{0.5}Sr_{1.5}Co_2Fe_2O_{22}$ hexaferrite samples sintered at 1050 ºC (top) and 1250 ºC. (bottom)

**Figure 5**. Temperature dependence of anisotropy field in $Ba_{0.5}Sr_{1.5}Co_2Fe_2O_{22}$ hexaferrite samples.

Figure 1

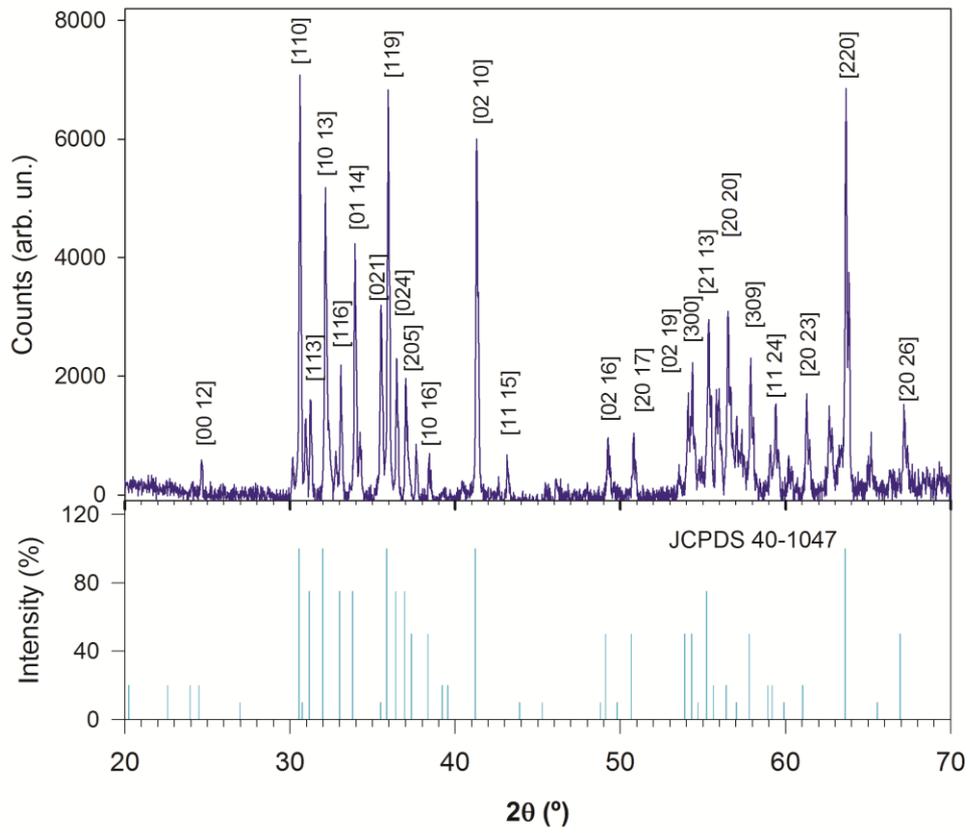

Figure 2

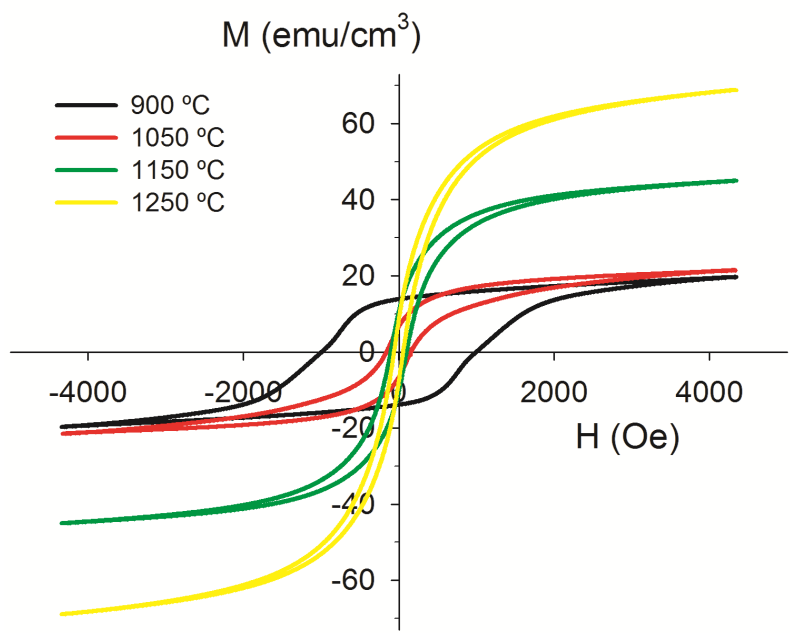

Figure 3

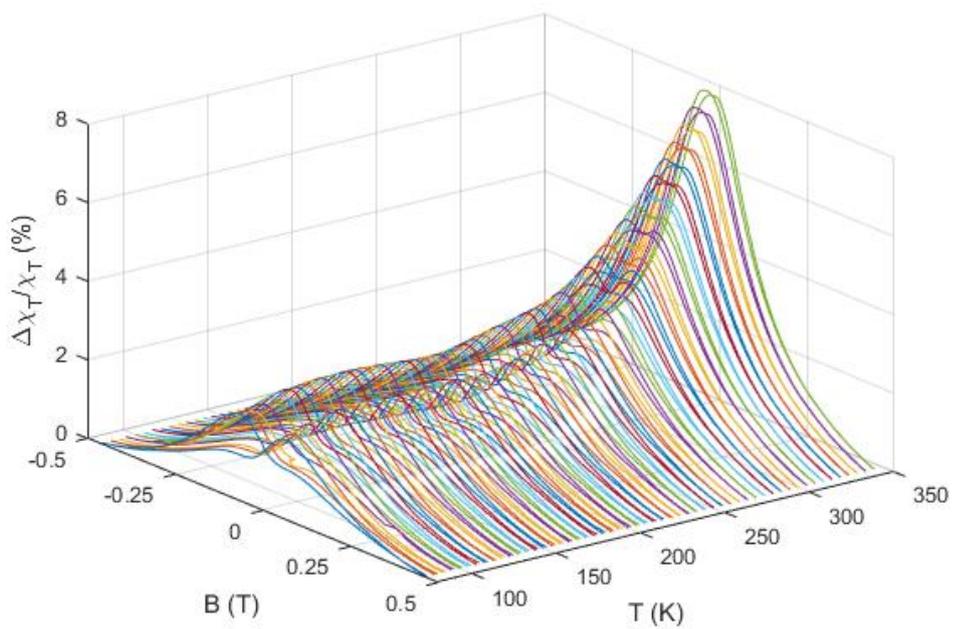

Figure 4

a)

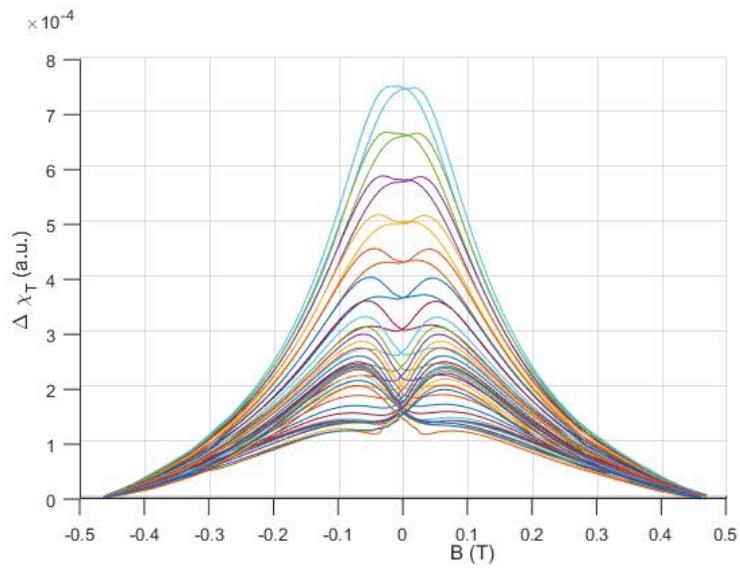

b)

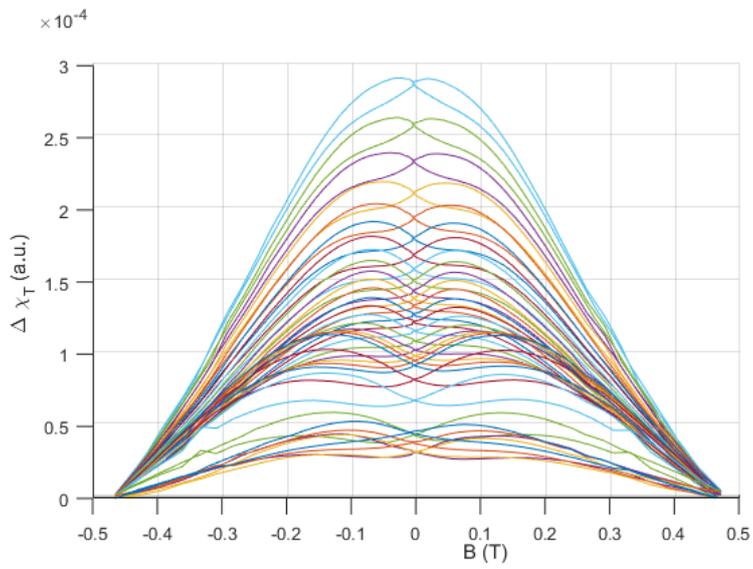

Figure 5

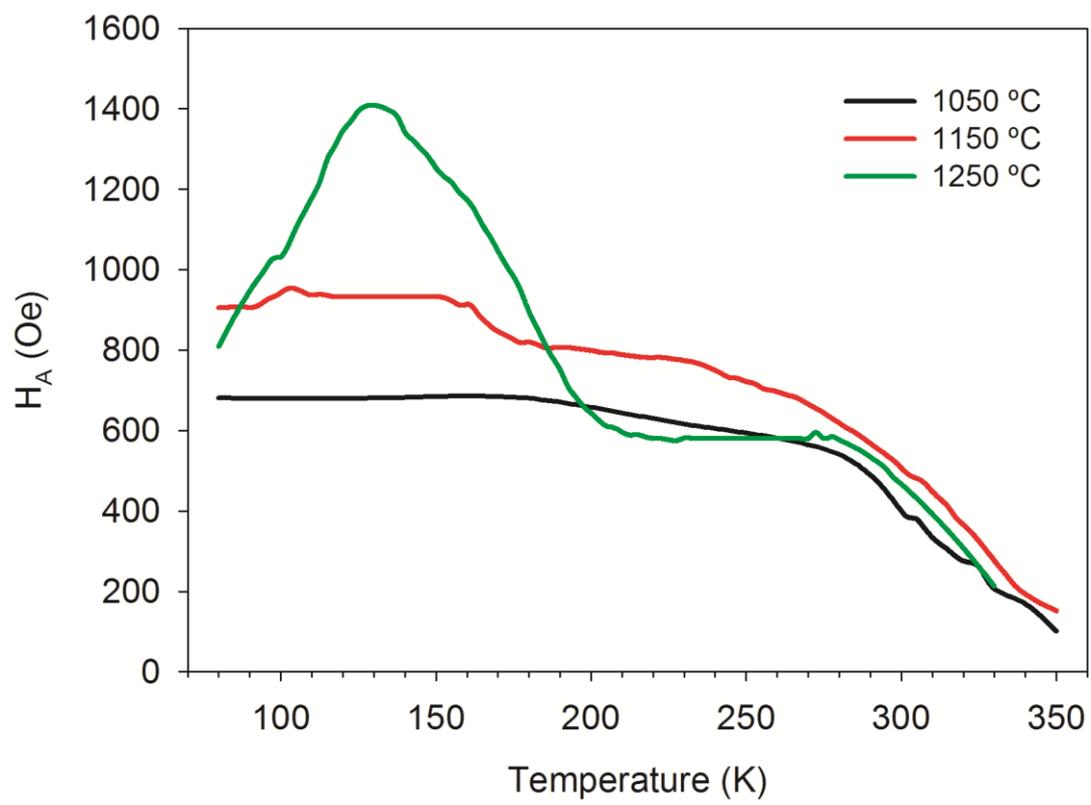